\newcommand{\docversion}{January 10, 2019}
\RequirePackage[l2tabu,orthodox,abort]{nag}
\documentclass{article}

\usepackage[utf8]{inputenc}
\usepackage{authblk}
\usepackage[font=small]{caption}
\usepackage{textcomp}
\usepackage[parskip,repro]{reidpr}

\title{Estimating influenza incidence using search query deceptiveness and generalized ridge regression}
\date{\docversion\ / LA-UR~18-24467}

\author[1,*]{Reid~Priedhorsky}
\author[1,2]{Ashlynn~R.~Daughton}
\author[3,$\dagger$]{Martha~Barnard}
\author[3,$\dagger$]{Fiona~O'Connell}
\author[1]{Dave Osthus}
\affil[1]{Los Alamos National Laboratory; Los Alamos, NM USA}
\affil[2]{University of Colorado, Boulder; Boulder, CO USA}
\affil[3]{Minnetonka Public Schools; Minnetonka, MN USA}
\affil[*]{Corresponding author: \url{reidpr@lanl.gov}}
\affil[$\dagger$]{These authors contributed equally to this work.}

\graphicspath{{./}{plot/}}

\pdfid{\docversion 7367c821-cfa7-4ebf-ae13-86a7df26e7c5}

\newcommand{\Figure}[1]{Fig~\ref{#1}}
\newcommand{\Table}[1]{Table~\ref{#1}}
\newcommand{\caphead}[1]{\textbf{#1.}}
\DeclareMathOperator*{\argmin}{arg\,min}

\newcommand{\suppILI}{S1~Dataset}
\newcommand{\suppGoogleFeatures}{S2~Dataset}
\newcommand{\suppOutput}{S3~Dataset}
\newcommand{\suppAllErrorFigures}{S4~Figure}

\begin{document}

\maketitle

%

\begin{abstract}

Seasonal influenza is a sometimes surprisingly impactful disease, causing thousands of deaths per year along with much additional morbidity.
Timely knowledge of the outbreak state is valuable for managing an effective response.
%
The current state of the art is to gather this knowledge using in-person patient contact.
While accurate, this is time-consuming and expensive.
This has motivated inquiry into new approaches using internet activity traces, based on the theory that lay observations of health status lead to informative features in internet data.

These approaches risk being deceived by activity traces having a coincidental, rather than informative, relationship to disease incidence; to our knowledge, this risk has not yet been quantitatively explored.
We evaluated both simulated and real activity traces of varying deceptiveness for influenza incidence estimation using linear regression.

We found that deceptiveness knowledge does reduce error in such estimates, that it may help automatically-selected features perform as well or better than features that require human curation, and that a semantic distance measure derived from the Wikipedia article category tree serves as a useful proxy for deceptiveness.
%
This suggests that disease incidence estimation models should incorporate not only data about how internet features map to incidence but also additional data to estimate feature deceptiveness.
%
By doing so, we may gain one more step along the path to accurate, reliable disease incidence estimation using internet data.
%
This capability would improve public health by decreasing the cost and increasing the timeliness of such estimates.

\end{abstract}

\section{Introduction}

Effective response to disease outbreaks depends on reliable estimates of their status.
This process of identifying new outbreaks and monitoring ongoing ones — \vocab{disease surveillance} — is a critical tool for policy makers and public health professionals~\cite{horstmann1974}.

The traditional practice of disease surveillance is based upon gathering information from in-person patient visits.
Clinicians make a diagnosis and report that diagnosis to the local health department.
These health departments aggregate the reports to produce local assessments and also pass information further up the government hierarchy to the national health ministry, which produces national assessments~\cite{mondor2012}.
Our previous work describes this process with a mathematical model~\cite{priedhorsky2018flow}.

This approach is accepted as sufficient for decision-making~\cite{johnson2004, rolfes2018annual} but is expensive, and results lag real time by anywhere from a week~\cite{cdc2016} to several months~\cite{bahk2015comparing, jajosky2004}.
Novel surveillance systems that use disease-related internet activity traces such as social media posts, web page views, and search queries are attractive because they would be faster and cheaper~\cite{priedhorsky2017cscw, santillana2015}.
One can conjecture that an increase of influenza-related web searches is due to an increase in flu observations by the public, which in turn corresponds to an increase in real influenza activity.
These systems use statistical techniques to estimate a mapping from past activity to past traditional surveillance data, then apply that map to current activity to predict current (but not yet available) traditional surveillance data, a process known as \vocab{nowcasting}.



\begin{figure}
  \includegraphics[width=\textwidth]{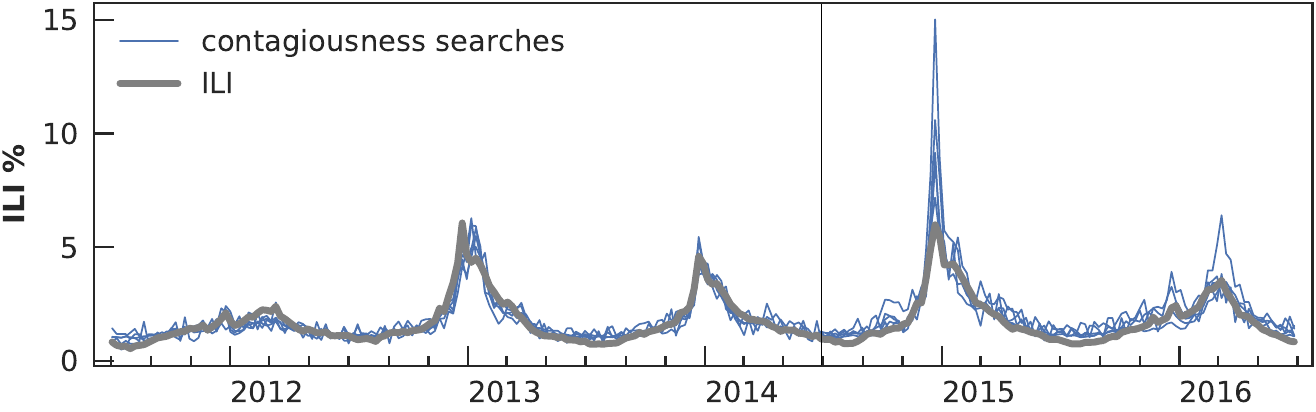}
  \caption{
    \caphead{Simple deceptive models for influenza-like illness (ILI)}
    This figure shows five one-feature models that map U.S.\ Google volume for contagiousness-related searches to ILI data from the CDC (``how long are you contagious'', ``flu contagious'', ``how long am i contagious'', ``influenza contagious'', ``when are you contagious'').
    We fit using ordinary least squares linear regression over the first three seasons of the study period.
    Despite a good fit during the training period, these models severely overestimate the peaks of the fourth and fifth seasons, demonstrating that a model's history of accuracy does not yield accurate predictions in the future if it uses deceptive input features, such as these contagiousness searches.
    Our \vocab{deceptiveness} metric quantifies the risk of such divergence.
  }
  \label{fig:us-ili-deceptive}
\end{figure}

One specific concern with this approach is that these models can learn coincidental, rather than informative, relationships~\cite{priedhorsky2018flow}.
For example, Bodnar and Salathé found a correlation between zombie-related Twitter messages and influenza~\cite{bodnar2013}.
More quantitatively, Ginsberg \etal\ built a flu model using search queries selected from 50~million candidates by a purely data-driven process that considered correlation with influenza-like-illness in nine regions of the United States~\cite{ginsberg2008}.
Of the 45 queries selected by the algorithm for inclusion in the model, 6~(13\%) were only weakly related to influenza (3~categorized as ``antibiotic medication'', 2~``symptoms of an unrelated disease'', 1~``related disease'').
Of the 55 next-highest-scoring candidates, 8~(15\%) were weakly related (3, 2, and 3 respectively) and 19~(35\%) were ``unrelated to influenza'', e.g. ``high school basketball''.
That is, even using a high-quality, very computationally expensive approach that leveraged demonstrated historical correlation in nine separate geographic settings, one-third of the top 100 features were weakly related or not related to the disease in question.
\Figure{fig:us-ili-deceptive} illustrates this problem for flu using contagiousness-related web searches.

Such features, with a dubious real link to the quantity of interest, pose a risk that the model may perform well during training but then provide erroneous estimates later when coincidental relationships fail, especially if they do so suddenly.
We have previously proposed a metric called \vocab{deceptiveness} to quantify this risk~\cite{priedhorsky2018flow}.
This metric quantifies the fraction of an estimate that depends on noise (coincidental relationships between input and output data) rather than signal (informative relationships) and is a real number between 0 and 1 inclusive.
We hypothesize that disease nowcasting models that leverage the deceptiveness of input features have better accuracy than those that do not.

This is an important question because disease forecasting is improved by better nowcasting.
For example, Brooks \etal's top-performing entry~\cite{brooks2018nonmechanistic} to the CDC's flu forecasting challenge~\cite{cdcchallenge2017} was improved by nowcasting.
Lu \etal\ tested autoregressive nowcasts using several internet data sources and found that they improved 1-week-ahead forecasts~\cite{lu2018accurate}.
Kandula \etal\ measured the benefit of nowcasting to their flu forecasting model at 8–35\%~\cite{kandula2018evaluation}.
Finally, our own work shows that a Bayesian seasonal flu forecasting model using ordinary differential equations benefits from filling a one-week reporting delay with internet-based nowcasts~\cite{osthus2018good}.

The present work tests this hypothesis using five seasons of influenza-like-illness (ILI) in the United States (2011–2016).
We selected U.S.\ influenza because high-quality reference data are easily available and because it is a pathogen of great interest to the public health community.
Although flu is often considered a mild infection, it can be quite dangerous for some populations, including older adults, children, and people with underlying health conditions.
Typical U.S.\ flu seasons kill ten to fifty thousand people annually~\cite{rolfes2018annual}.

Our experiment is a simulation study followed by validation using real data.
This lets us test our approach using fully-known deceptiveness as well as a more realistic setting with estimated deceptiveness.
We trained linear estimation models to nowcast ILI using an extension of ridge regression~\cite{hoerl1970} called \vocab{generalized ridge}~\cite{hemmerle1975gridge} or \vocab{gridge regression} that lets us apply individual weights to each feature, thus expressing a prior belief on their value: higher value for lower deceptiveness.
We used three classes of input features:
(1)~synthetic features constructed by adding plausible noise patterns to ILI,
(2)~Google search volume on query strings related to influenza,
and (3)~Google search volume on inferred topics related to influenza.

We found that accurate deceptiveness knowledge did indeed reduce prediction error, and in the case of the automatically-generated query string features, as much or more than topic features that require human curation. We also found that semantic distance as derived from the Wikipedia article category tree served as a useful proxy for deceptiveness.

The remainder of this paper is organized as follows.
We next describe our data sources, regression approach, and experiment structure.
After that, we describe our results and close with their implications and suggestions for future work.

\section{Methods}

Our study period was five consecutive flu seasons, 2011–2012 through 2015–2016, using weekly data.
The first week in the study was the week starting Sunday, July 3, 2011, and the last week started Sunday, June 26, 2016, for a total of 261 consecutive weeks.
We considered each season to start on the first Sunday in July, and the previous season to end on the day before (Saturday).

We used gridge regression to fit input features to U.S.\ ILI over a subset of the first three seasons, then used the fitted coefficients to estimate ILI in the fourth and fifth seasons.
(We used this training schedule, rather than training a new model for each week as one might do operationally, in order to provide a more challenging setting to better differentiate the models.)
We assessed accuracy by comparing the estimates to ILI using three metrics: $r^2$ (the square of Pearson correlation), root mean squared error (RMSE), and hit rate.

The experiment is a full factorial experiment with four factors, yielding a total of 225 models:
\begin{enumerate}

  \item \inhead{Class of input features} (3 levels): synthetic features, search query string volume, and search topic volume.

  \item \inhead{Training period} (3): one, two, or three consecutive seasons.

  \item \inhead{Noise added to deceptiveness} (5): perfect knowledge of deceptiveness to none at all.

  \item \inhead{Model type} (5): ridge regression and four levels of gridge regression.

\end{enumerate}

This procedure is implemented in a Python program.

The remainder of this section describes our data sources, regression algorithm, experimental factors, and assessment metrics in detail.

\subsection{Data sources}


We used four types of data in this experiment:
\begin{enumerate}

  \item \inhead{Reference data.}
    U.S.\ national ILI from the Centers for Disease Control and Prevention (CDC).
    This is a weekly estimate of influenza incidence.

  \item \inhead{Synthetic features.}
    Weekly time series computed by adding specific types of systematic and random noise to ILI.
    These simulated features have known deceptiveness.

  \item \inhead{Flu-related concepts.}
    We used the crowdsourced Wikipedia category tree to enumerate a set of concepts and estimate the semantic relatedness of each to influenza.

  \item \inhead{Real features.}
    Two types of weekly time series: Google search query strings and Google search topics.
    These features are based on the flu-related concepts above and use estimated flu relatedness as a proxy for deceptiveness.

\end{enumerate}
This section explains the format and quirks of the data, how we obtained them, and how readers can also obtain them.

\subsubsection{Reference data: U.S.\ influenza-like illness (ILI) from CDC}


\vocab{Influenza-like illness} (ILI) is a syndromic metric that estimates influenza \vocab{incidence}, i.e., the number of new flu infections.
It is the fraction of patients presenting to the health care system who have symptoms consistent with flu and no alternate explanation~\cite{cdc2016}.
The basic process is that certain clinics called \vocab{sentinel providers} report the total number of patients seen during each interval along with those diagnosed with ILI.
Local, provincial, and national governments then collate these data to produce corresponding ILI reports~\cite{cdc2016}.

U.S.\ ILI values tend to range between 1–2\% during the summer and 7–8\% during a severe flu season~\cite{cdc2016, cdc2018percentage}.
While an imperfect measure (for example, it is subject to reporting and behavior biases if some groups, like children, are more commonly seen for ILI~\cite{lee2015detecting}), it is considered sufficiently accurate for decision-making purposes by the public health community~\cite{johnson2004, rolfes2018annual}.

In this study, we used weekly U.S.\ national ILI downloaded from the Centers for Disease Control and Prevention (CDC)'s FluView website~\cite{cdc2017fluview} on December 21, 2016, six months after the end of the study period.
This delay is enough for reporting backfill to settle sufficiently~\cite{osthus2018good}.
\Figure{fig:us-ili-deceptive} illustrates these data, and they are available as an Excel spreadsheet in \suppILI\ (\code{ILI.xls}).

\subsubsection{Synthetic features: Computed by us}
\label{sec:features-synthetic}

These simulated features are intended to model a plausible way in which internet activity traces with varying usefulness might arise.
Their purpose is to provide an experimental setting with features that are sufficiently realistic and have known deceptiveness.

\begin{figure}
  \includegraphics[width=\textwidth]{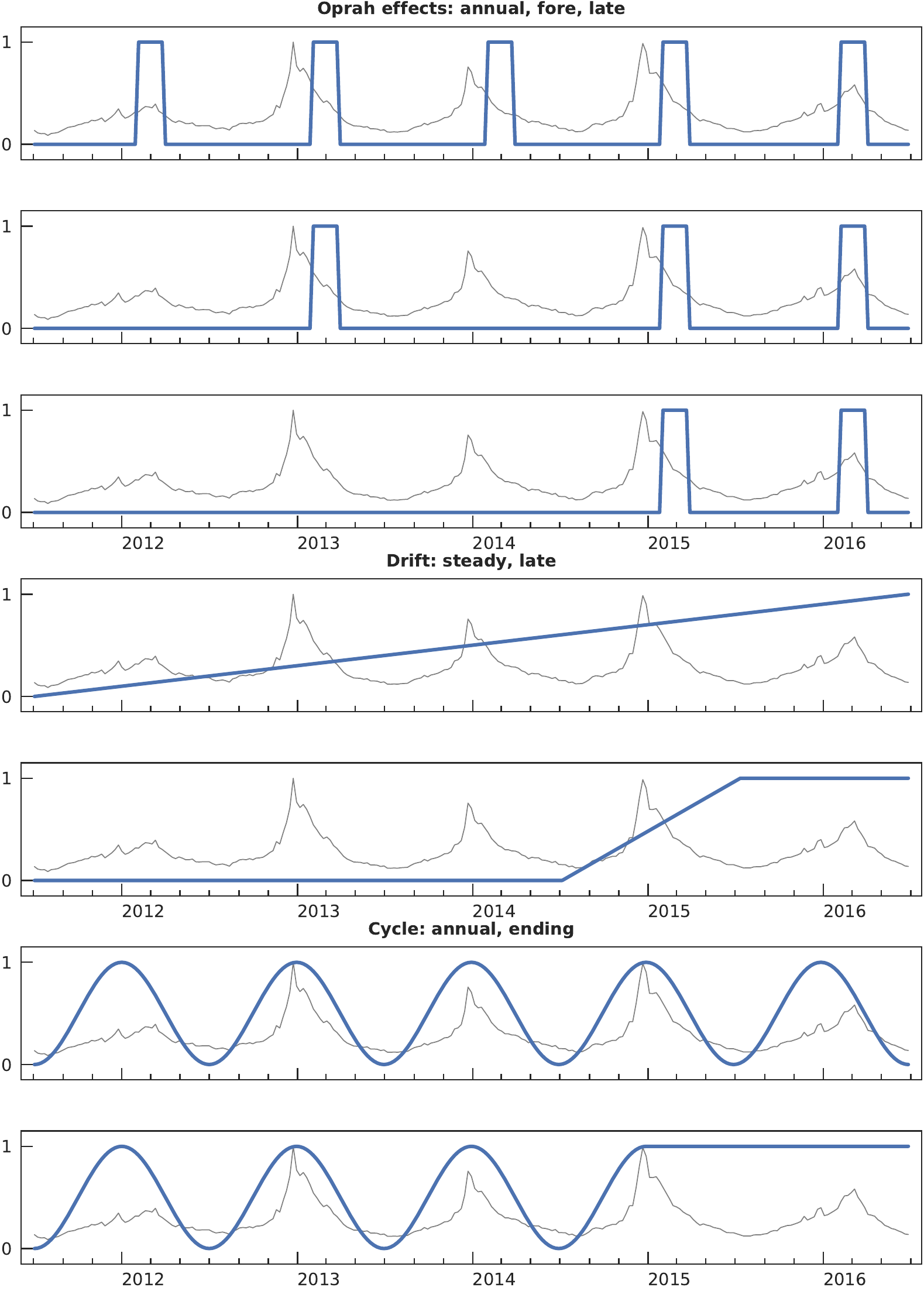}
  \caption{
    \caphead{Systematic noise basis functions for synthetic features}
    These functions model plausible mechanisms of how features are affected by exogenous events.
  }
  \label{fig:bases}
\end{figure}

Each synthetic feature $x$ is a random linear combination of ILI $y$, Gaussian random noise $\varepsilon$, and systematic noise $\sigma$ (all vectors with 261 elements, one for each week):
\begin{align}
  x &= w_\text{i} y + w_\text{r} \varepsilon + w_\text{s} \sigma \\
  \label{eq:three-1}
  1 &= w_\text{i} + w_\text{r} + w_\text{s}
\end{align}
A feature's deceptiveness $g \in [0,1]$ is simply the weight of its systematic noise: $g = w_\text{s}$.

Random noise $\varepsilon$ is a random multivariate normal vector with standard deviation 1.
Systematic noise $\sigma$ is a random linear combination of seven basis functions $\sigma_j$:
\begin{align}
  w_\text{s} \sigma &= \sum_{j=1}^7 w_{\text{s}j} \sigma_j \\
         w_\text{s} &= \sum_{j=1}^7 w_{\text{s}j}
\end{align}
These basis functions, illustrated in \Figure{fig:bases}, simulate sources of systematic noise for internet activity traces.
They fall into three classes:
%
\begin{itemize}

  \item \inhead{Oprah effect~\cite{priedhorsky2018flow}:} 3 types.
    These simulate pulses of short-lived traffic driven by media interest.
    For example, U.S.\ Google searches for measles were 10 times higher in early 2015 than any other time during the past five years~\cite{google2017trends}, but measles incidence peaked in 2014~\cite{cdc2018measles}.


    The three specific bases are: \vocab{annual}~$\sigma_1$, a pulse every year shortly after the flu season peak; \vocab{fore}~$\sigma_2$, pulses during both the training and test seasons (second, fourth, and fifth); and \vocab{late}~$\sigma_3$, pulses only during the test seasons (fourth and fifth).
    The last creates features with novel divergence after training is complete, producing deceptive features that cannot be detected by correlation with reference data.

  \item \inhead{Drift:} 2 types.
    These simulate steadily changing public interest.
    For example, as the case definition of autism was modified, the number of individuals diagnosed with autism increased~\cite{hill2015}.

    The two bases are: \vocab{steady}~$\sigma_4$, a slow change over the entire study period of five seasons, and \vocab{late}~$\sigma_5$, a transition from one steady state to another over the fourth season.
    The latter again models novel divergence.

    %
    %

  \item \inhead{Cycle:} 2 types.
    This simulates phenomena that have an annual ebb and flow correlating with the flu season.
    An example is the U.S.\ basketball season noted above.

    The two bases are: \vocab{annual}~$\sigma_6$, cycles continuing for all five seasons, and \vocab{ending}~$\sigma_7$, cycles that end after the training seasons.
    The latter again models novel divergence.

\end{itemize}

In order to build one feature, we need to sample the nine elements of the weight vector $w = ( w_\text{i}, w_\text{r}, w_{\text{s}1}, ...\, w_{\text{s}7} )$, which sums to 1.
This three-step procedure is as follows.
First, $w_\text{i}$, $w_\text{r}$, and $w_\text{s}$ are sampled from a Dirichlet distribution:
%
\begin{gather}
  (w_\text{i}, w_\text{r}, w_\text{s}) \sim \text{Dir}(1.0, 0.5, 1.5) \\
  \text{E}(w_\text{i}) = \frac{1}{3} \quad , \quad
  \text{E}(w_\text{r}) = \frac{1}{6} \quad , \quad
  \text{E}(w_\text{s}) = \frac{1}{2}
\end{gather}
Next, the relative weight of the three types of systematic noise is sampled:
\begin{align}
        (w_\text{so}, w_\text{sd}, w_\text{sc})
  &\sim w_\text{s} \text{Dir}(0.3,0.3,0.3) 
\end{align}
Finally, all the weight for each type is randomly assigned with equal probability to a single basis function:
\begin{align}
     (w_{\text{s}1}, w_{\text{s}2}, w_{\text{s}3})
  &\sim w_\text{so} \text{Multinomial}(1, [\frac{1}{3}, \frac{1}{3}, \frac{1}{3}]) \\
     (w_{\text{s}4}, w_{\text{s}5})
  &\sim w_\text{sd} \text{Multinomial}(1, [\frac{1}{2}, \frac{1}{2}]) \\
     (w_{\text{s}6}, w_{\text{s}7})
  &\sim w_\text{sc} \text{Multinomial}(1, [\frac{1}{2}, \frac{1}{2}]) \\
\end{align}

\begin{figure}
  \includegraphics[width=\textwidth]{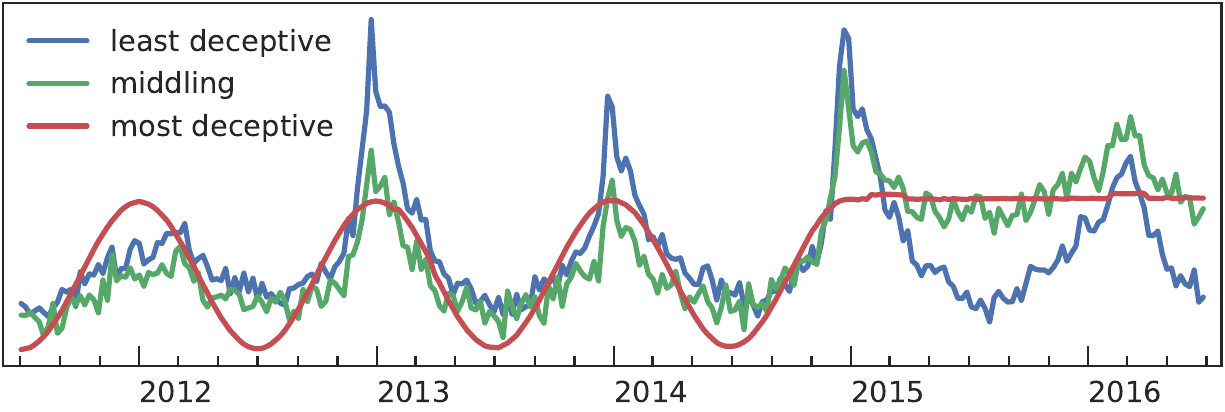}
  \caption{
    \caphead{Example synthetic features}
    This figure shows the least and most deceptive features and a third with medium deceptiveness.
  }
  \label{fig:synthetic-examples}
\end{figure}

\begin{figure}
  \includegraphics[width=\textwidth]{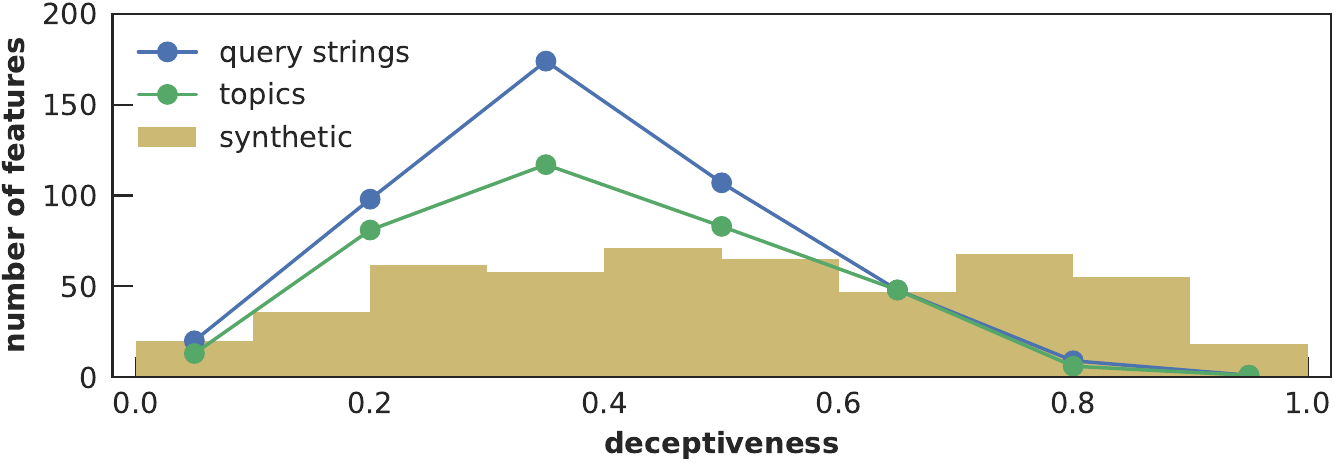}
  \caption{
    \caphead{Deceptiveness histogram for all three feature types}
    Synthetic features have continuous deceptiveness and are grouped into ten bins, while the two types of search features have seven discrete deceptiveness levels.
  }
  \label{fig:decept-all}
\end{figure}

This procedure yields a variety of high- and low-quality synthetic features.
We sampled 500 of them.
\Figure{fig:synthetic-examples} shows three examples, and \Figure{fig:decept-all} shows the deceptiveness histogram.
The features we used are available in \suppOutput\ (\code{synthetic.xlsx}).

\subsubsection{Flu-related concepts: Wikipedia}
\label{sec:flu-concepts}

In order to build the flu nowcasting model based on web search queries described in the next section, we first needed a set of influenza-related concepts.
We used the Wikipedia inter-article link graph to generate a set of candidate concepts and the Wikipedia article category hierarchy to estimate the semantic relatedness of each concept to influenza, which we use as a proxy for deceptiveness.
An important advantage of this approach is that it is automated and easily generalizable to other diseases.

Wikipedia is a popular web-based encyclopedia whose article content and metadata are crowdsourced~\cite{ayers2008}.
We used two types of metadata from a dataset~\cite{priedhorsky2017cscwdata} collected March 24, 2016 for our previous work~\cite{priedhorsky2017cscw} using the Wikipedia API.

First, Wikipedia articles contain many hyperlinks to other articles within the encyclopedia.
This work used the article ``Influenza'' and the 572 others it links to, including clearly related articles such as ``Infectious disease'' and apparently unrelated ones such as ``George W.\ Bush'', who was the U.S.\ president immediately prior to 2009 H1N1.

Second, Wikipedia articles are leaves in a category hierarchy.
Both articles and categories have one or more parent categories.
For example, one path from ``Influenza'' to the top of the tree is: \emph{Healthcare-associated infections}, \emph{Infectious diseases}, \emph{Diseases and disorders}, \emph{Health}, \emph{Main topic classifications}, \emph{Articles}, and finally \emph{Contents}.
This tree can be used to estimate semantic relatedness between two articles.
The number of levels one must climb the tree before finding a common category is a metric called \vocab{category distance}~\cite{priedhorsky2017cscw}; the distance between an article and itself is 1.
For example, the immediate categories of ``Infection'' include \emph{Infectious diseases}.
Thus, the distance between these two articles is~2, because we had to ascend two levels from ``Influenza'' before discovering the common category \emph{Infectious diseases}.

We used category distance between each of the 573 articles and ``Influenza'' to estimate the semantic relatedness to influenza.
The minimum category distance was 1 and the maximum 7.

The basic intuition for this approach is that Wikipedia category distance is a reasonable proxy for how related a concept is to influenza, and this relatedness is in turn a reasonable proxy for deceptiveness.
For example, consider a distance-1 feature and a distance-7 feature that are both highly correlated with ILI.
Standard linear regression will give equal weight to both features.
However, we conjecture that the distance-7 feature's correlation is more likely to be spurious than the distance-1's; i.e., we posit that the distance-7 feature is more deceptive.
Thus, we give the distance-1 feature more weight in the regression, as described below.

Because category distance is a discrete variable $d \in [1,7] \cap \mathbb{Z}$, while deceptiveness $g \in [0,1]$ is continuous, we convert category distance into a deceptiveness estimate $\hat g$ as follows:
\begin{equation}
  \hat g = (1-2\epsilon) \frac{d-1}{6} + \epsilon
\end{equation}
The purpose of $\epsilon \ne 0$ is to ensure that features with minimum category distance of~1 receive regularization from the linear regression, as described below.
In our initial data exploration, the value of $\epsilon$ had little effect, so we used $\epsilon = 0.05$; therefore, $\hat g \in \{ 0.05, 0.20, 0.35, 0.50, 0.65, 0.80, 0.95 \}$.
We emphasize that category distance is already a noisy proxy for deceptiveness.
Even with zero noise added to deceptiveness, $\hat g \ne g$.

It is important to realize that because Wikipedia is continually edited, metadata such as links and categories change over time.
Generally, mature topic areas such as influenza and infectious disease are more stable than, for example, current events.
The present study assumes that the dataset we used is sufficiently correct despite its age; i.e., freshly collected links and categories might be somewhat different but not necessarily more correct.

The articles used and their category distances are in \suppGoogleFeatures\ (\code{en+Influenza.xlsx}).

\subsubsection{Real features: Google searches}
\label{sec:features-search}


Typically, each feature for internet-based disease surveillance estimates public interest in a specific concept.
This study uses Google search volume as a measure of public interest.
By mapping our Wikipedia-derived concepts to Google search queries, we obtained a set of queries with estimated deceptiveness.
Then, search volume over time for each of these queries, as well as their deceptiveness, are input for our algorithms.

We tested two types of Google searches.
\vocab{Search query strings} are the raw strings typed into the Google search box.
We designed an automated procedure to generate query strings from Wikipedia article titles.
\vocab{Search topics} are concepts assigned to searches by Google using proprietary and unspecific algorithms.
We built a map from Wikipedia article titles to topics manually.

Our procedure to map articles to query strings is:
\begin{enumerate}
  \item Decode the percent notation in the title portion of the article's URL~\cite{wikipedia2018percent}.
  \item Change underscores to spaces.
  \item Remove parentheticals.
  \item Approximate non-ASCII characters with ASCII using the Python package \code{unidecode}~\cite{solc2018}.
  \item Change upper case letters to lower case. (This serves a simplifying rather than necessary purpose, as the Google Trends API is case-insensitive.)
  \item Remove all characters other than alphanumeric, slash, single quote, space, and hyphen.
  \item
    Remove \vocab{stop phrases} we felt were unlikely to be typed into a search box.
    Matches for the following regular expressions were removed (note leading and trailing spaces):
    \begin{itemize}
    \item ``\code{ and\b}''
    \item ``\code{^global }''
    \item ``\code{^influenza .? virus subtype }''
    \item ``\code{^list of }''
    \item ``\code{^the }''
    \end{itemize}
\end{enumerate}
This produces a query string for all 573 articles.
The map is 1-to-1: each article maps to exactly one query string, and each query string maps to exactly one article.
The process is entirely automated once the list of stop phrases is developed.

Google search topics is a somewhat more amorphous concept.
Searches are assigned to topics by Google's proprietary machine learning algorithms, which are not publically available~\cite{google2018trendshelp}.
A given topic is identified by its name or a hexadecimal code.
For example, the query string ``apple'' might be assigned to ``Apple (fruit)'' or ``Apple (technology company)'' based on the content of the full search session or other factors.

\begin{table}
  \makebox[\textwidth][c]{
  \begin{tabular}{lllc}
    \toprule
      \textbf{Article}
    & \textbf{Query string}
    & \textbf{Topic name}
    & \textbf{Topic code} \\
    \midrule
    Sense (molecular biology)           & sense                               & Sense (Molecular biology)              & \code{/m/0dpw95} \\
    Influenza A virus subtype H9N2      & h9n2                                & Influenza A virus subtype H9N2 (Virus) & \code{/m/0b3dc1} \\
    George W.\ Bush                     & george w bush                       & George W. Bush (43rd U.S. President)   & \code{/m/09b6zr} \\
    \bottomrule
  \end{tabular}
  }
  \caption{
    \caphead{Sample Wikipedia articles and the queries to which they map}
  }
  \label{tab:sample-queries}
\end{table}

To manually build a mapping between Wikipedia articles and Google search topics,
we entered the article title and some variations into the search box on the Google Trends website~\cite{google2017trends} and then selected the most reasonable topic named in the site's auto-complete box.
The topic code was in the URL.
If the appropriate topic was unclear, we discussed it among the team.
Not all articles had a matching topic; we identified 363 topics for the 573 articles (63\%).
Among these 363 articles, the map is 1-to-1.

\Table{tab:sample-queries} shows a few examples of both mappings, and \Figure{fig:decept-all} shows the deceptiveness histograms.

\begin{figure}
  \includegraphics[width=\textwidth]{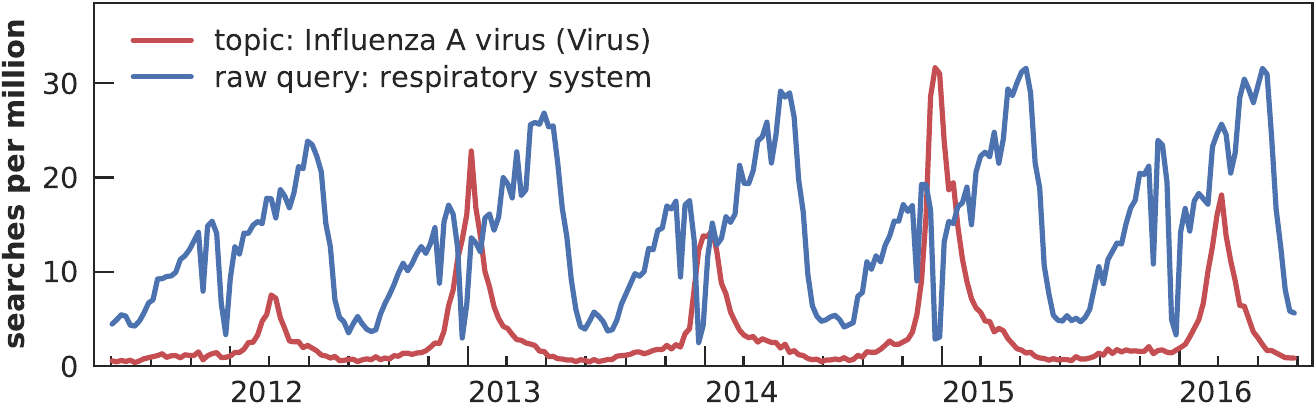}
  \caption{
    \caphead{Search volume for two queries presented to our model}
    The topic ``Influenza A virus (Virus)'' shows a seasonal pattern roughly corresponding to ILI, while the raw query ``respiratory system'' shows a seasonal pattern that does not correspond to ILI.
  }
  \label{fig:ght-examples}
\end{figure}

We downloaded search volume for both query strings and topics from the Google Health Trends API~\cite{stocking2017} on July 31, 2017.
This gave us a weekly time series for the United States for each search query string and topic described above.
These data appear to be a random sample, changing slightly from download to download.

Each element of the time series is the probability that the reported search session contains the string or topic, multiplied by 10 million.
Searches without enough volume to exceed an unspecified privacy threshold are set to zero, i.e., we cannot distinguish between few searches and no searches.
For this reason, we removed from analysis searches with more than 2/3 of the 261 weeks having a value of zero.
This resulted in 457 usable of 573 query strings (80\%) and 349 of 363 topics (96\%).
\Figure{fig:ght-examples} shows two of these time series.

Our map and category distances are in \suppGoogleFeatures\ (\code{en+Influenza.xlsx}).
Google's terms of service prohibit redistribution of the search volume data.
However, others can request the data using the same procedure we used~\cite{google2018ghtform}.
We used the script \code{ght_get} in the experiment source code for downloading.
This script depends on a patched source code file originally provided by Google that has an unclear license; therefore, we cannot redistribute this code.
Access to the Google source code is granted with the data, and we do provide the patch.

\subsection{Gridge regression}

Linear regression is a popular approach for mapping features (inputs) to observations (output).
This section describes the algorithm and the extensions we used in our experiment to incorporate deceptiveness information.

The model for linear regression is
\begin{equation}
  y = X\beta + \varepsilon
\end{equation}
where $y$ is an $N \times 1$ observation vector, $X = [1, x_1, x_2, \ldots, x_p]$ is an $N \times (p+1)$ feature matrix where $x_i$ is the $N \times 1$ standardized feature vector (i.e., mean centered and standard deviation scaled) corresponding to feature $i$, $\beta$ is a $(p+1) \times 1$ coefficient vector, $\varepsilon \sim \text{MVN}(0,\sigma^2 I)$, where $\text{MVN}(\mu,\Sigma)$ is a multivariate normal distribution with mean $\mu$ and covariance matrix $\Sigma$, $\sigma^2 > 0$ is a scalar, and $I$ is an $N \times N$ identity.
Standardizing the features of $X$ is a convention that places all features on the same scale.

The goal of linear regression is to find the estimate of $\beta$ that minimizes the sum-of-squared residual errors. The \vocab{ordinary least squares} (OLS) estimator $\hat{\beta}^\text{OLS}$ solves the following:
\begin{equation}
  \label{eq:leastsquares}
    \hat{\beta}^{\text{OLS}}
  = \argmin_{\beta}
            \sum_{j=1}^N \Bigg(y_j - \beta_0 - \sum_{i=1}^p \beta_i x_{ji}\Bigg)^2
\end{equation}
or in matrix form:
\begin{equation}
    \hat{\beta}^{\text{OLS}}
  = (X'X)^{-1} Xy
\end{equation}
%

$\hat{\beta}^\text{OLS}$ is the unbiased, minimum variance estimator of $\beta$, assuming $\varepsilon$ is normally distributed~\cite[ch.~1–2]{scheffe1959variance}.
A prediction corresponding to a new feature vector $\ddot{x}$ is $\hat{y}^\text{OLS} = \ddot{x} \hat{\beta}^\text{OLS}$.

While $\hat{y}^\text{OLS}$ is unbiased, it is often possible to construct an estimator with smaller \vocab{expected prediction error} — i.e., with predictions on average closer to the true value — by introducing some amount of bias through \vocab{regularization}, which is the process of introducing additional information beyond the data.
Also, regularization can make regression work in situations with more features than observations, like ours.

One popular regularization method is called \vocab{ridge regression}~\cite{hoerl1970}, which extends OLS by encouraging the coefficients $\beta$ to be small.
This minimizes:
\begin{equation}
  \label{eq:ridge_min}
    \hat{\beta}^{\text{ridge}}_{\lambda}
  = \argmin_{\beta} \left\{
        \sum_{j=1}^N
          \left( y_j - \beta_0 - \sum_{i=1}^p \beta_i x_{ji} \right)^{\!\!2}
      + \lambda \sum_{i=1}^p \beta_i^2
    \right\}
\end{equation}
or equivalently:
\begin{equation}
  \label{eq:ridge_closed_form}
    \hat{\beta}^{\text{ridge}}_{\lambda}
  = (X'X + \lambda I )^{-1} X'y.
\end{equation}
The additional parameter $\lambda \geq 0$ controls the strength of regularization. When $\lambda=0$, this is equivalent to OLS.
As $\lambda$ increases, the coefficient vector $\beta$ is constrained towards zero more vociferously.
Ridge regression applies the same degree of regularization to each feature, as $\lambda$ is common to all features.


A second extension, called \vocab{generalized ridge regression}~\cite{hemmerle1975gridge} or \vocab{gridge regression}, adds a feature-specific modifier $\kappa_i$ to the regularization:
\begin{equation}
  \label{eq:fridge_min}
    \hat{\beta}^{\text{gridge}}_{\lambda\kappa}
  = \argmin_{\beta} \left\{
        \sum_{j=1}^N
          \left( y_j - \beta_0 - \sum_{i=1}^p \beta_i x_{ji} \right)^{\!\!2}
      + \lambda \sum_{i=1}^p \kappa_i \beta_i^2
    \right\}
\end{equation}
$\kappa_i \geq 0$ adjusts the regularization penalty individually for each feature (ridge regression is a special case where $\kappa_i = 1\ \forall\ i$).
Gridge retains closed-form solvability:
\begin{equation}
  \label{eq:gridge_closed_form}
    \hat{\beta}^{\text{gridge}}_{\lambda\kappa}
  = (X'X + \lambda K)^{-1} X'y
\end{equation}
where $K$ is a diagonal matrix with $\kappa_i$ on the diagonal and zero on the off-diagonals.

Gridge regression allows us to incorporate feature-specific deceptiveness information by making $\kappa_i$ a function of feature $i$'s deceptiveness.
The more deceptive feature $i$, the larger $\kappa_i$.

\subsection{Experiment factors}

Our experiment had 225 conditions.
This section describes its factors:
input feature class~(3 levels),
training period~(3),
deceptiveness noise added~(5),
and regression type~(5).

\subsubsection{Input feature class}

We tested three classes of input features:
\begin{enumerate}

  \item \inhead{Synthetic.} Randomly generated transformations of ILI, as described above in~\S\ref{sec:features-synthetic}.

  \item \inhead{Search query string.} Volume of Google searches entered directly by users, as described above in \S\ref{sec:features-search}.

  \item \inhead{Search topic.} Volume of Google search topics inferred by Google's proprietary algorithms, as described above in \S\ref{sec:features-search}.

\end{enumerate}
Each feature comprises a time series of weekly data, with frequency and alignment matching our ILI data.


\subsubsection{Training period}

We tested three different training periods:
1st through 3rd seasons inclusive (three season),
2nd and 3rd (two seasons),
and 3rd only (one season).
Because the 4th season contains transitions in the synthetic features, we did not use it for training even when testing on the 5th season.

\subsubsection{Deceptiveness noise added}

The primary goal of our study is to evaluate how much knowledge of feature deceptiveness helps disease incidence models.
In the real world, this knowledge will be imperfect.
Thus, one of our experiment factors is to vary the quality of feature deceptiveness knowledge.

Our basic approach is to add varying amounts of noise to the best available estimate of each feature's deceptiveness $\hat g \in [0,1]$.
Recall that for synthetic features, $\hat g = g$ is known exactly, while for the search-based features, $\hat g$ is an estimate based on the Wikipedia category distance.

To compute the noise-added deceptiveness $\tilde g_i$ for feature $i$, for noise added $\gamma$, we simply select a random other feature $j$ and mix with its deceptiveness: $\tilde g_i = (1-\gamma) \hat g_i + \gamma \hat g_j$.
There are five levels of this factor:
\begin{itemize}
  \item Zero noise: $\gamma = 0$, i.e., the model gets the best available estimate of $g_i$.
  \item Low noise: $\gamma = 0.05$.
  \item Medium noise: $\gamma = 0.15$.
  \item High noise: $\gamma = 0.4$.
  \item Total noise: $\gamma = 1$, i.e., the model gets no correct information at all about $g_i$.
\end{itemize}
Models do not know what condition they are in; they get only $\tilde g_i$, not $\gamma$.

\subsubsection{Regression type}

We tested five types of gridge regression:
\begin{enumerate}

  \item Ridge regression: $\kappa_i = 1$, i.e., ignore deceptiveness information.

  \item
    Threshold gridge regression: keep features with category distance $d_i \leq 3$ and discard them otherwise, as in \cite{priedhorsky2017cscw}.
    This is implemented as a threshold $\tilde g_i = 0.35$, which is applicable to both search and synthetic features (which have no $d_i$).
    \begin{equation}
      \kappa_i = \begin{cases}
                    0.1 & \text{if } \tilde g_i \leq 0.35 \\
                      1 & \text{otherwise}
                 \end{cases}
    \end{equation}

  \item Linear fridge: $\kappa_i = \tilde g_i$.

  \item Quadratic fridge: $\kappa_i = \tilde g_i^2$.

  \item Quartic fridge: $\kappa_i = \tilde g_i^4$.

\end{enumerate}
These levels are in rough ascending order of deceptiveness importance.
(We additionally tested, but do not report, a few straw-man models to help identify bugs in our code.)

All models used $\lambda = 150.9$, obtained by 10-fold cross-validation~\cite{james2013introduction}.
For each model, we tested 41 values of $\lambda$ evenly log-spaced between $10^{-1}$ and $10^7$; each fold fitted a model on the 9 folds left in and then evaluated its RMSE on the one fold left out.
The $\lambda$ with the lowest mean RMSE (plus a bias of up to 0.02 to encourage $\lambda$s in the middle of the range) across the 10 folds, was reported as the best $\lambda$ for that model.
We then used the mean of these best $\lambda$s for our experiment.

\subsection{Assessment of models}

To evaluate a model, we apply its coefficients learned during the training period to input features during the 52 weeks of the fourth and fifth seasons respectively, yielding estimated ILI $\hat y$.
We then compare $\hat y$ to reference ILI $y$ for each of the two test seasons.
For each model and metric, this yields two scalars.

We report three metrics:
\begin{enumerate}

  \item \inhead{$\boldsymbol{r^2}$},
    the square of the Pearson correlation $r$.
    Most previous work reports this unitless metric.

  \item \inhead{Root mean squared error} (RMSE),
    defined as:
    \begin{equation}
      \sqrt{\frac{1}{N} \sum_{j=1}^N (y_j - \hat{y}_j)^2}
    \end{equation}
    has interpretable units of ILI.

  \item \inhead{Hit rate}
    is a measure of how well a prediction captures the direction of change.
    It is defined as the fraction of weeks when the direction of the prediction (increase or decrease) matches the direction of the reference data~\cite{santillana2015}:
    \begin{equation}
      \frac{\sum_{j=2}^N \text{sign}(y_j - y_{j-1})
            \overset{?}{=}
            \text{sign}(\hat y_j - \hat y_{j-1} )}
           {N-1}
    \end{equation}
    Because it captures the trend (is the flu going up or down?), it directly answers a simple, relevant, and practical public health question.



\end{enumerate}

\section{Results}

\begin{figure}
  \includegraphics[width=\textwidth]{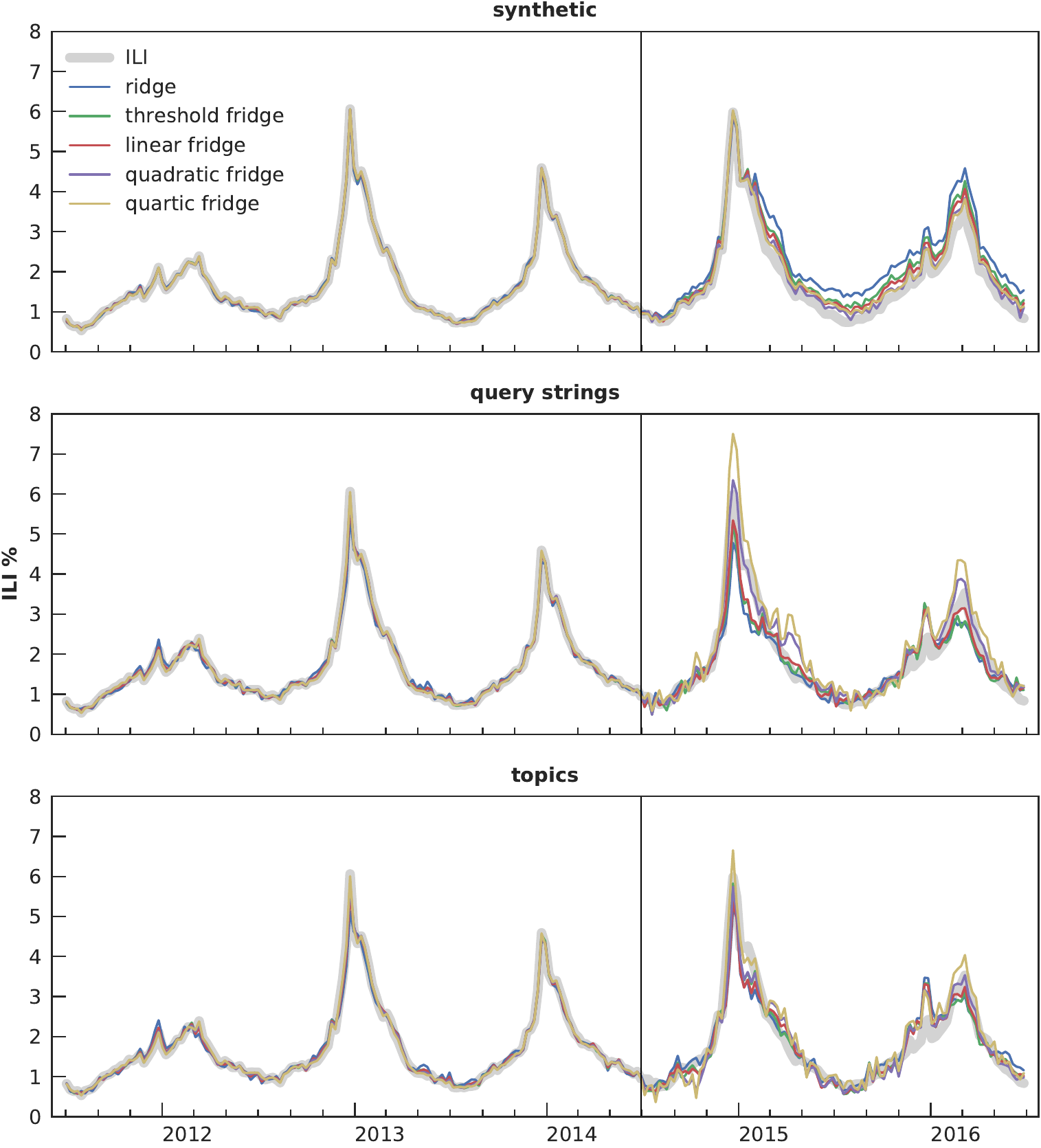}
  \caption{
    \caphead{ILI predictions for zero-added-noise, 3-season-trained models}
    This figure shows the 15 models in the ``ideal'' situation: no noise added to deceptiveness, and trained on all three training seasons.
    The different types of gridge regression show subtle yet distinct differences, with the models taking into account deceptiveness more generally being closer to the ILI reference data.
  }
  \label{fig:selected-models}
\end{figure}

Output of our regression models is illustrated in \Figure{fig:selected-models}, which shows 15 selected conditions.
These conditions are close to what would be done in practice: use all available training information and add no noise.
The differences between gridge regression types are subtle, but they are real, and close examination shows that the stronger gridge models that place higher importance on deceptiveness information are closer to the ILI reference data.
The remainder of this section analyzes these differences across all the conditions.

\begin{figure}
  \includegraphics[width=\textwidth]{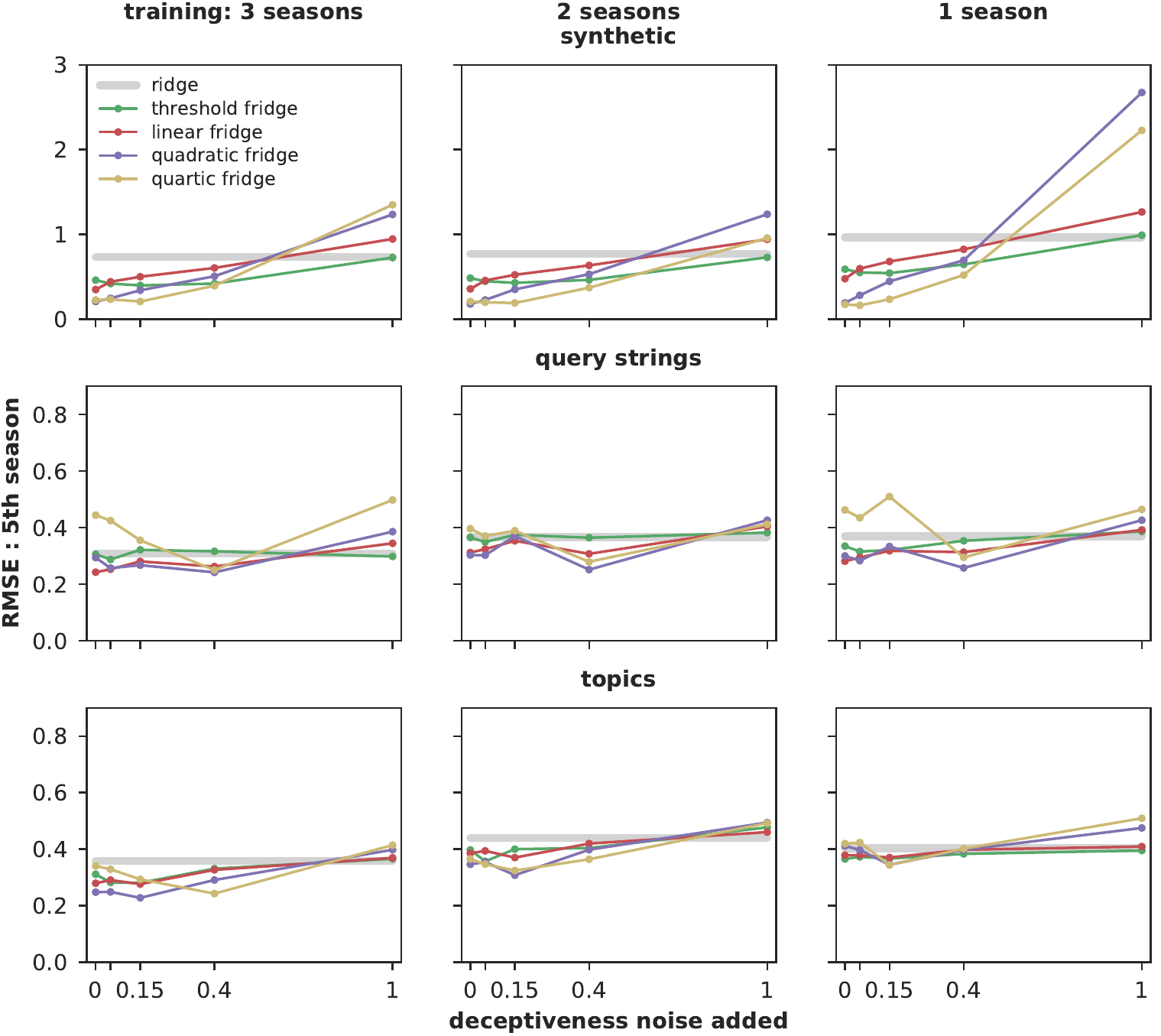}
  \caption{
    \caphead{Gridge regression error compared to plain ridge, 5th season}
    Gridge algorithms that take into account deceptiveness \textit{usually} have lower RMSE than plain ridge, which does not.
    Further, adding deceptiveness noise usually gives the appropriate trend: worse deceptiveness knowledge means worse predictions.
    $r^2$ and RMSE on the 4th season show similar trends, while hit rate shows limited benefit from gridge; these figures are available in \suppAllErrorFigures.
  }
  \label{fig:results-rmse}
\end{figure}

\Figure{fig:results-rmse} illustrates the effect on error (RMSE) of adding noise to deceptiveness information; $r^2$ is similar.
Generally, the gridge algorithms have lower error than plain ridge in lower-added-noise conditions and higher error in higher-added-noise situations.
That is, conditions with better knowledge of deceptiveness outperform the baseline, and performance declines as deceptiveness knowledge worsens, which is the expected trend.
This supports our hypotheses that (a)~incorporating knowledge of feature deceptiveness can improve estimates of disease incidence based on internet data and (b)~semantic distance, as expressed in the Wikipedia article category tree, is an effective proxy for deceptiveness.
(Hit rate shows limited benefit for gridge, as we discuss below.)


\begin{figure}
  \vspace{-24pt}
  \includegraphics[width=\textwidth]{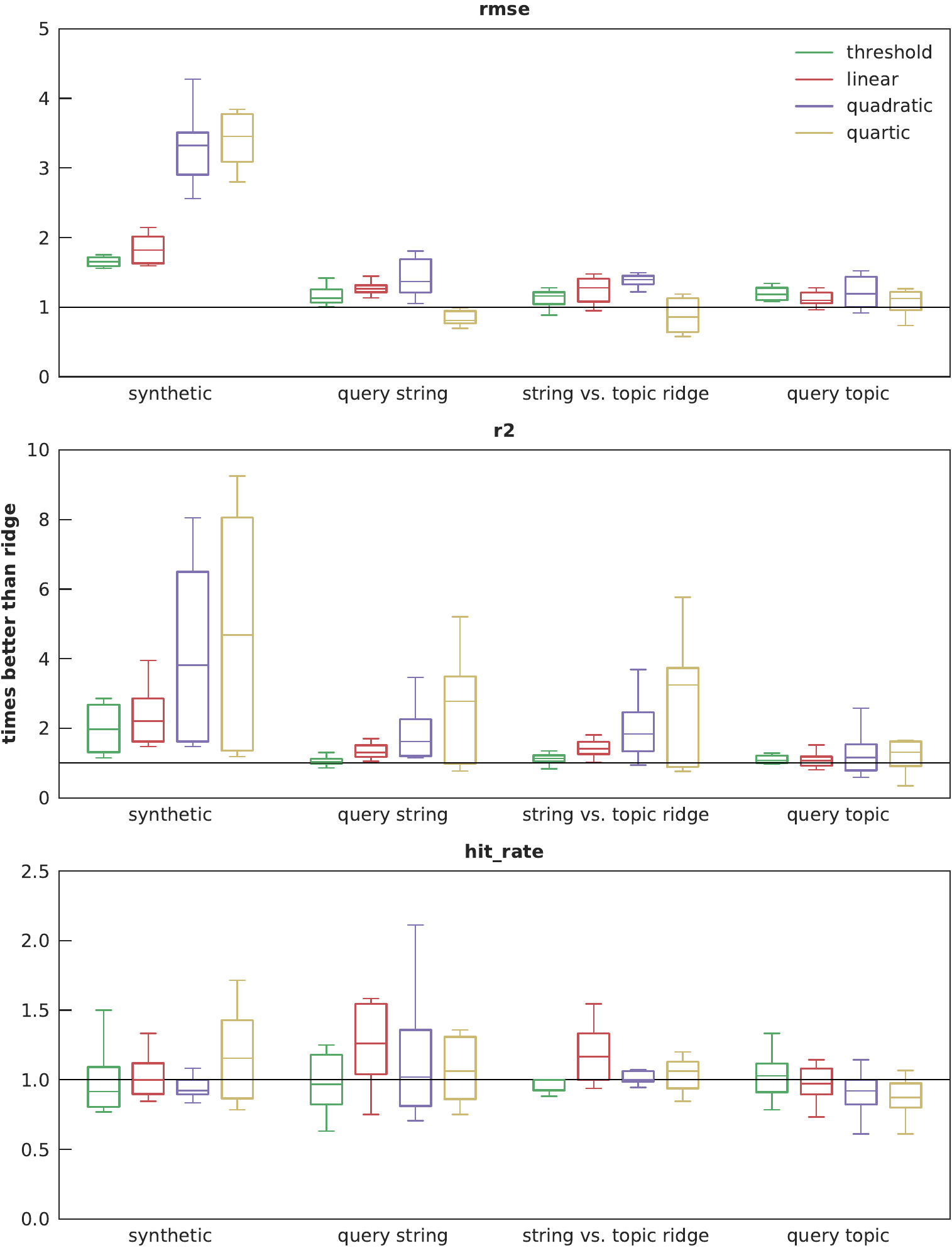}
  \caption{
    \caphead{Improvement over ridge in zero- and low-noise conditions}
    This figure illustrates performance of the gridge algorithms divided by plain ridge in the same condition.
    The third group compares \emph{query string} gridge models against \emph{topic} ridge.
    Boxes plot the median with first and third quartiles, and the whiskers show the maximum and minimum.
    Each box-and-whiskers summarizes 12 data points.
    For the two error metrics, increasing the importance of deceptiveness through quadratic gridge yields increasing improvement over plain ridge, but the trend ends at quartic gridge.
    Notably, this is true even when comparing query string models (which are completely automatic) against topic ridge (which requires lots of manual attention), suggesting that deceptiveness information can be used to replace expensive human judgement.
    Hit rate, however, seems to gain limited benefit from deceptiveness in this experiment.
  }
  \label{fig:results-summary}
\end{figure}

\Figure{fig:results-summary} summarizes the improvement of the four gridge algorithms over plain ridge for all three metrics in the zero- and low-noise conditions.
These results suggest that adding low-noise knowledge of deceptiveness to ridge regression improves error but not hit rate.
It also appears that the benefits of gridge level off between quadratic (deceptiveness squared) and quartic (deceptiveness raised to the fourth power): while quartic sometimes beats quadratic ridge, it frequently is worse than plain ridge.

We speculate that the lack of observed benefit of gridge on hit rate is due to one or both of two reasons.
First, plain ridge may be sufficiently good on this metric that it has already reached diminishing returns; recall that in \Figure{fig:selected-models} all five algorithms captured the overal trend of ILI well.
Second, ILI is noisy, with lots of ups and downs from week to week regardless of the medium-term trend.
This randomness may limit the ability of hit rate to assess performance on a weekly time scale without overfitting.
That is, we believe that gridge's failure to improve over plain ridge on hit rate is unlikely to represent a concerning flaw in the algorithm.

\section{Discussion}

Our previous work introduced \vocab{deceptiveness}, which quantifies the risk that a disease estimation model's error will increase in the future because it uses features that are coincidentally, rather than informatively, correlated with the quantity of interest~\cite{priedhorsky2018flow}.
This work tests the hypothesis that incorporating deceptiveness knowledge into a disease nowcasting algorithm reduces error; to our knowledge, it is the first work to quantitatively assess this question.
To do so, we used simulated features with known deceptiveness as well as two types of real web search features with deceptiveness estimated using semi- and fully-automated algorithms.

\subsection{Findings}

Our experiment yielded three main findings:

\begin{enumerate}

\item
  Deceptiveness information does help our linear regression nowcasting algorithms, and it helps more when it is more accurate.

\item
  A readily available, crowdsourced semantic relatedness measure, Wikipedia category distance, is a useful proxy for deceptiveness.

\item
  Deceptiveness information helps automatically generated features perform the same or better than similar, semi-automated features that require human curation.

\end{enumerate}

The effects we measured are stronger for the synthetic features than the real ones.
We speculate that this is for two reasons.
First, the web search feature types are skewed towards low deceptiveness, because they are based on Wikipedia articles directly linked from ``Influenza'', while the synthetic features lack this skew.
Second, the synthetic features can have zero-noise deceptiveness information, while the real features cannot, because they use Wikipedia category distance as a less-accurate proxy.
If verified, the second would further support the hypothesis that more accurate deceptiveness information improves nowcasts.

The third finding is interesting because it is relevant to a long-standing tension regarding how much human intervention is required for accurate measurements of the real world using internet data: more automated algorithms are much cheaper, but they risk oversimplifying the complexity of human experience.
For example, our query strings were automatically generated from Wikipedia article titles, which are written for technical accuracy rather than salience for search queries entered by laypeople.
To select features for disease estimation, one could use a fully-automated approach (e.g., our query strings), a semi-automated approach (e.g., our topics, which required a manual mapping step), or a fully-manual approach (e.g., by expert elicitation of search queries or topics, which we did not test).

One might expect that a trade-off would be present here: more automatic is cheaper, but more manual is more accurate.
However, this was not the case in our results.
The third box plot group in \Figure{fig:results-summary} compares the gridge models using query string features to a baseline of plain ridge on \emph{topic} features.
Query strings perform favorably regardless of whether the baseline is plain ridge on query strings or topics, and sometimes the improvement is greater than gridge using topic features.
This suggests that there is not really a trade-off, and fully automatic features might be among the most accurate.

\subsection{Limitations}

All experiments are imperfect.
Due to its finite scope, this work has many limitations.
We believe that the most important ones are:

\begin{itemize}

\item
  Wikipedia is changing continuously.
  While we believe that these changes would not have a material effect on our results, we have not tested this.

\item
  Wikipedia has non-semantic categories, such as the roughly 20,000 articles in category \emph{Good article} that our algorithm that would assign distance~1 from each other.
  We have not yet encountered any other relevant non-semantic categories, and ``Influenza'' is not a \emph{Good article}, so we believe this limitation does not affect the present results.
  However, any future work extending our algorithms should exclude these categories from the category distance computation.

\item
  The mapping from Wikipedia articles to Google query strings and topics has not been optimized.
  While we have presented mapping algorithms that are reasonable both by inspection and supported by our current and prior~\cite{priedhorsky2017cscw} results, we have not compared these algorithms to alternatives.

\item
  Linear regression algorithm metaparameters were not fully evaluated.
  For example, $\epsilon$ in §\ref{sec:flu-concepts} was thoughtfully but arbitrarily assigned rather than experimentally optimized.

\item
  Other methods of feature generation may be better.
  This experiment was not designed to evaluate the full range of feature generation algorithms.
  In particular, direct elicitation of features such as query strings and topics should be evaluated.

\end{itemize}

\subsection{Future work}

This is an initial feasibility study using a fairly basic nowcasting model.
At this point, the notion of deceptiveness for internet-based disease estimation is promising, but continued and broader positive results are needed to be confident in this hypothesis.
In addition to addressing the limitations above, we have two groups of recommendations for future work.

First, multiple opportunities to improve nowcasting performance should be investigated.
Additional deceptiveness-aware fitting algorithms such as generalized lasso~\cite{tibshirani2011glasso} and generalized elastic net~\cite{sokolov2016glasticnet} should be tested.
Category distance also has opportunities for improvement.
For example, it can be made finer-grained by measuring path length through the Wikipedia category tree rather than counting the number of levels ascended: the distance between ``Influenza'' and ``Infection'' would become~3, taking into account that \emph{Infectious diseases} was a direct category of the latter.
Finally, alternate deceptiveness estimates need testing, for example category distance based on the medical literature.
In addition to better nowcasting, utility needs to be demonstrated when deceptiveness-aware nowcasts augment best-in-class forecasting models, such as those doing well in the CDC's flu forecasting challenge~\cite{cdcchallenge2017}.

Second, we are optimistic that our algorithms will generalize well to different diseases and locations.
This is because our best feature-generation algorithm is fully automated, making it straightforward to generalize by simply offering new input.
For example, to generate features for dengue fever in Colombia, one could: start with the article ``Dengue fever'' in Spanish Wikipedia; write a set of Spanish stop phrases; use the Wikipedia API to collect links from that article and walk the category tree; pull appropriate search volume data from Google or elsewhere; and then proceed as described above.
Future studies should evaluate generalizability to a variety of disease and location contexts.

\subsection{Conclusion}

We present a study testing the value of deceptiveness information for nowcasting disease incidence using simulated and real internet features and generalized ridge regression. We found that incorporating such information does in fact help nowcasting; to our knowledge, ours is the first quantitative evaluation of this question.

Based on these results, we hypothesize that other internet-based disease estimation methods may also benefit from including feature deceptiveness estimates. We look forward to further research yielding deeper insight into the deceptiveness question.

\section*{Author contributions}

Experiment concept: RP~ARD~DO.
Methods: RP~ARD~DO (experiment), MB~FO (mapping Wikipedia articles to Google searches), DO (statistical approach).
Data curation, investigation, and software: RP~ARD.
Visualization: RP.
Literature review: RP~ARD.
Writing: RP~ARD~DO.
Review and editing: RP~ARD~MB~FO~DO.
Funding acquisition and project administration: RP.

\section*{Acknowledgments}

We thank the Wikipedia editing community for building the link and category networks used to compute semantic distance and Google, Inc.\ for providing us with search volume data.
We also appreciate the helpful feedback provided by anonymous reviewers.

\bibliographystyle{plainurl}
\bibliography{refs}

\end{document}